\documentclass[useAMS,usenatbib]{mnras}

\usepackage{newtxtext,newtxmath}

\usepackage[T1]{fontenc}

\DeclareRobustCommand{\VAN}[3]{#2}
\let\VANthebibliography\thebibliography
\def\thebibliography{\DeclareRobustCommand{\VAN}[3]{##3}\VANthebibliography}

\usepackage{graphicx}	
\usepackage{amsmath}	
\usepackage{amssymb}	
\usepackage{longtable}

\title[The mass, spin and rotational energy of the RBHs]{The mass, spin and rotational energy of the remnant black holes from compact binary mergers}
\author[Can-Min Deng]{Can-Min Deng  \thanks{E-mail: dengcm@ustc.edu.cn} \\
Department of Astronomy, School of Physical Sciences, University of Science and Technology of China, Hefei, 230026, Anhui, China\\
CAS Key Laboratory for Research in Galaxies and Cosmology, Department of Astronomy, University of Science and Technology of China, Hefei 230026, Anhui, China\\
}

\date{Accepted XXX. Received YYY; in original form 8 Mar 2020}

\pubyear{2020}

\begin{document}
\label{firstpage}
\pagerange{\pageref{firstpage}--\pageref{lastpage}}
\maketitle




\begin{abstract}
Recently, many gravitational wave events from compact binary mergers have been detected by LIGO. Determining the final mass and spin of the remnant black holes (RBHs) is a fundamental issue and is also important in astrophysics. In this paper, unified models for predicting the final mass and spin of the RBHs from compact binary mergers is proposed. The models achieve a good accuracy within the parameter range of interest. In addition, the rotational energy of the RBHs is also studied which is relevant to the electromagnetic counterparts of the mergers. It is found the distribution of the rotational energy of the RBHs from different types of mergers of compact binary has its own characteristics, which might help identify the electromagnetic counterparts associated with the mergers.

\end{abstract}

\section{Introduction}
The compact binary mergers are the most promising sources of the gravitational waves that can be detected by LIGO. Since LIGO detected the first gravitational wave event GW150914 \citep{2016PhRvL.116f1102A}, a total of eleven compact binary mergers were detected during the O1 and O2 \citep{2019PhRvX...9c1040A}. Ten of them were binary black hole (BH) mergers. One of them was a double neutron star (NS) merger, which was the first detection of a double NS merger event \citep{2017PhRvL.119p1101A}. Recently, O3 is in operation, and candidates of BH-NS mergers were also detected (GCN 25324, 25695, 25814, 25876, 26350).  Also, a candidate binary NS merger event GW190425 is reported \citep{2020ApJ...892L...3A}. When NS-NS merges, such as GW170817, a stable remnant NS is possible only if the equation of state (EOS) of the NS is stiff enough \citep{2018ApJ...861..114Y,2018ApJ...860...57A}. 
And the properties of remnant black holes (RBHs) from compact binary mergers which is what the spin and mass is is a fundamental question.
There are a lot of numerical simulations in the literature that have looked at this question, such as simulations for binary BH merger in references \cite{2007PhRvD..76f4034B,2017PhRvD..95f4024J}, NS-BH merger in references \citep{2010PhRvD..82d4049K,2011PhRvD..84f4018K,2015PhRvD..92d4028K} and binary NS merger in references \cite{2013PhRvD..87b4001H,2015PhRvD..91l4041D,2017PhRvD..95b4029D}.
The mass of the RBH is not much different from the initial total mass of the system, and the mass loss is basically a few percent.
The spin of the RBH, however, is very complex and depends on the parameters of the binary, such as the mass ratio, the initial spin and the EOS of the NS.
Based on the numerical simulation results, models have been proposed to predict the final mass and spin from the merger of binary BHs \citep{2007PhRvD..76f4034B,2016ApJ...825L..19H,2016PhRvD..93d4006H,2009ApJ...704L..40B,2008PhRvD..78h1501T,2008PhRvD..78d4002R,2008ApJ...674L..29R,2008PhRvD..77b6004B,2017PhRvD..95f4024J,2019PhRvD..99f4045V}. Model of predicting the final mass and spin of the RBH from the BH-NS mergers have also been studied \citep{2013PhRvD..88j4025P}. These models can achieve good accuracy \citep{2013PhRvD..88j4025P,2017PhRvD..95f4024J,2019PhRvD..99f4045V}, but it is complex and not very convenient to apply for astrophysical interests. For NS-NS merger, note that \cite{2019MNRAS.489L..91C} gave formulas for estimating the final mass and spin. And such formulas are very useful for multimessenger parameter estimation \citep{2019EPJA...55...50R,2019MNRAS.489L..91C,2020PhRvD.101j3002K,2020arXiv200211355D}.
For these reasons, I attempt to seek a unified model with good accuracy  to estimate the final mass and spin of the RBHs for compact binaries mergers.

As exciting as the first detection of a gravitational wave signal, the electromagnetic counterparts of the GW170817 were observed, namely the short GRB 170817A and the kilonova AT2017gfo \citep{2017ApJ...848L..12A}. In fact, when the first binary BH merger gravitational wave event GW150914 was detected, a suspected short GRB counterpart was found \citep{2016ApJ...826L...6C,2016ApJ...827L..38G}.  Recently, a candidate optical counterpart to the binary BH merger gravitational wave event S190521g
is reported by \cite{2020arXiv200614122G}.
In addition, a BH-NS merger candidates have also been found that it might be associated with a short gamma-ray signal GBM-190816 \citep{2019arXiv191200375Y}. 
With the successful detection of gravitational waves and their electromagnetic counterparts, astronomy ushered in the era of gravitational waves and multi-messenger. The observational search and theoretical study on the electromagnetic counterparts of gravitational waves from compact binary coalecences is one of the most hottest topics currently.  As we known, short GRB and mergernova/kilonova are the most likely electromagnetic counterparts for compact binary mergers \citep{2019arXiv191205659N}.  
The popular central engine model for short GRB is that the merger leaves behind a BH and accretion disk which amplifies the magnetic fields to extract the BH's rotational energy through the BZ mechanism \citep{1977MNRAS.179..433B}, producing relativistic jets that produce GRB itself and its afterglows \citep{2015PhR...561....1K}. 
In addition, the extended emission and the X-ray plateau in short GRBs are also possibly related to the extraction of the rotational energy of the RBH \citep{2013ApJ...768...63L,2015ApJ...804L..16K,2017ApJ...846..142K}. 
In the case of double NS merger, if the remnant is a BH, then the brightness and light curve of the corresponding mergernova will be different with when the remnant is a massive NS \citep{2013ApJ...771...86G,2013ApJ...776L..40Y,2017ApJ...837...50G,2018ApJ...852L...5M}.
Even for double BHs mergers, there might also be the associated short GRBs \citep{2016ApJ...821L..18P}. 
And all of these are relevant to the extraction of the rotational energy of the RBH.
The rotational energy of the RBH depends on the parameters of the binary, however,  which is remained to be studied in detail. Therefore, the second task of this paper is to systematically explore the distribution of the rotational energy of the RBHs with different parameters .

\section{Model for estimating the mass and spin of the remnant BH}
Considering such compact stars binary, one star has a mass of $m_1$ and the other star has a mass of $m_2$ ($m_2\geq m_1$), and $M=m_{1}+m_{2}$. The corresponding dimensionless spins are $a_1$ and $a_2$ respectively. Here, define the mass ratio as $q=m_2/m_1$.

After the binary reaches the innermost stable orbit (ISCO), the final plunge will occur.
The dimensionless orbital angular momentum of the binary at the ISCO is given by 
\begin{equation}
\hat{L}_{\mathrm{ISCO}}=A_{a} \frac{m_{1} m_{2}}{M^{2}} = A_{a} \eta,
\label{eq:LISCO}
\end{equation}
where $\eta=q/(1+q)^2$ and  $A_{a}$ is a constant.  In the small mass ratio limit and for nonspinning bodies,  $A_{a}=2 \sqrt{3}\simeq 3.464$ \citep{2007PhRvD..76f4034B}.

 If the two stars have their own spin, then the  initial total dimensionless spin angular momentum is
\begin{equation}
\chi= \frac{a_{1} m_{1}^{2}+a_{2} m_{2}^{2}}{\left(m_{1}+m_{2}\right)^{2}}= \frac{\left(a_{1}+a_{2} q^{2}\right)}{(1+q)^{2}}.
\label{eq:chi}
\end{equation}
Here  only  the nonprocessing case is considered where the orbital angular momentum and the spin angular momentum are parallel or antiparallel.

In the process of merger, the angular momentum of the system will undergo extremely complicated evolution. We hope that the spin
of the RBH can be obtained by using a simple parameterized model instead of the complex numerical relativistic calculation.
According to equation (\ref{eq:LISCO}) and (\ref{eq:chi}), the binary system has an initial total angular momentum, before the final plunge, $J=a_{0}M^{2}$, where $a_{0}=A_{a}  \eta + B_{a} \chi$, $A_{a}$  and $B_{a} $ are constants  to be determined. In the subsequent merger process, gravitational wave radiation will radiate away a small part of the mass and take away a small part of the angular momentum.
In the cases of BH-NS or NS-NS mergers, after the NS is tidally disrupted, some material will be ejected, and it also takes away a small amount of mass and angular momentum. Then during the merging process, the change of the total angular momentum $dJ=M^{2}da+2aMdM$. Using a small quantity approximation, we have $\Delta a =\Delta J/M^{2}-2a\Delta M/M$. Then the final spin of the RBH $a_{\rm{f}}=a_{0}+\Delta a=a_{0}(1-2\Delta M/M)+\Delta J/M^{2} \simeq a_{0}+\Delta J/M^{2} $, where $\Delta M/M \ll 1 $. Motivated by the fact that $\Delta J/M^{2} \propto \Delta M/M$ is established in the case of  BH-BH merger \citep{2007PhRvD..76f4034B}, we finally 
obtain the  approximate model  for estimating the $a_{\rm{f}}$, that is 
\begin{equation}
a_{\rm{f}}=A_{a} \eta+B_{a} \chi-C_{a}\left(1-m_{\rm{f}}\right),
\label{eq:af}
\end{equation}
where $m_{\rm{f}}=M_{\rm{f}}/M=(M-\Delta M)/M$ is the dimensionless final mass of the RBH.  $C_{a}$ is also a constant  to be determined. 
Although the above model is motivated by the case of BH-BH merger, I will generalize it to the case containing NSs, and one will see satisfactory results. Next, let's explore the model for estimating the $m_{\rm{f}}$.

To estimating the $m_{\rm{f}}$, I adopt a phenomenological model. Again for the case of  non-spinning BH-BH merger, we have $(1-m_{\rm{f}}) \propto \eta^{2}$ \citep{2007PhRvD..76f4034B}. On the other hand, I find that, with an increasing $\chi$,  $m_{\rm{f}}$ increases almost linearly. Therefore let's adopt the following model to estimate $m_{\rm{f}}$ ,
\begin{equation}
m_{\rm{f}}(\rm{BH-BH})=A_{\rm{m}} +B_{\rm{m}} \eta^{2}+C_{\rm{m}} \chi,
\label{eq:BH-BH}
\end{equation}
for BH-BH merger. $A_{\rm{m}}$, $B_{\rm{m}}$  and $C_{\rm{m}}$ are unknown constants to be determined.
However, one would expects that $A_{\rm{m}} \sim 1$, $B_{\rm{m}}$ and $C_{\rm{m}}$  are negative values.

For BH-NS merger, the equation of state (EOS) of the NS will play a role, because the NS may be tidally disruptied during the merger. Hence adding a term related to the tidal deformability $\Lambda$ of the NS to the model phenomenologically, that is  
\begin{equation}
m_{\rm{f}}(\rm{BH-NS})=A_{\rm{m}} +B_{\rm{m}} \eta^{2}+C_{\rm{m}} \chi+D_{\rm{m}} \log{\Lambda}.
\label{eq:BH-NS}
\end{equation}
where $A_{\rm{m}}$,  $B_{\rm{m}}$, $C_{\rm{m}}$ and $D_{\rm{m}}$ are the unknown constants to be determined.

For NS-NS merger,I find that the total mass of the binary also play an important role according to the data. Ignoring the spin of the NSs, one has
\begin{equation}
m_{\rm{f}}(\rm{NS-NS})=A_{\rm{m}} +B_{\rm{m}} \eta^{2}+C_{\rm{m}} \log{M}+D_{\rm{m}} \log{\tilde{{\Lambda}}}
\label{eq:NS-NS}
\end{equation}
where $A_{\rm{m}}$,  $B_{\rm{m}}$, $C_{\rm{m}}$ and $D_{\rm{m}}$ are also the unknown constants to be determined. Here I use the rescale tidal deformability $\tilde{{\Lambda}}$ for the two NSs in the binary.

In this work, I restrict our study to non-spinning NSs due to the numerical simulation data are based on non-spinning NSs. For a typical pulsar with rotation
period ${P_{{\rm{NS}}}} \sim 10\,{\rm{ms}}$, the  nondimensional spin parameter
${a_{{\rm{NS}}}} \sim 0.004{\kappa _{ - 1}}P_{{\rm{NS}},10{\rm{ms}}}^{ - 1}$, where $\kappa$ is an parameter that describes how
centrally condensed the NS is, which depends on the EOS \citep{2011PhRvD..83l4035L}. Therefore it is reasonable to assume $a_{\rm{NS}}=0$.

\section{Model Fitting}
\subsection{Data Selection}
For BH-BH merger,  the results of numerical relativity presented in reference \cite{2017PhRvD..95f4024J} (see its table XIV) are adopted. 25 sets of data $(q,{a_1},{a_2},m_{\rm{f}},a_{\rm{f}})$  were included in \cite{2017PhRvD..95f4024J}, where only the data with $q\leq4$, $-0.85 \leq a_1 \leq 0.85$, $-0.85 \leq a_2 \leq 0.85$ and ${a_{\rm{f}}}>0.1$ are considered for the purposes of this paper. The data are fitted using equation (\ref{eq:af}) and (\ref{eq:BH-BH}) .

For BH-NS merger, the results of numerical relativity when the mass of NS ${M_{{\rm{NS}}}} = 1.35{M_ \odot }$ from references \cite{2010PhRvD..82d4049K,2011PhRvD..84f4018K,2015PhRvD..92d4028K} are adopted.  Here, we have ${M_1} = {M_{{\rm{NS}}}}$ and ${M_2} = {M_{{\rm{BH}}}}$.  It should be pointed out that the general assumption in numerical simulation is that $a_{\rm{NS}}=0$. 
Then $\chi$ in the equation (\ref{eq:af}) degenerates to ${a^{{\rm{BH}}}_{{\rm{0}}}}{q^2}/{(1 + q)^2}$ for BH-NS merger, where ${a^{{\rm{BH}}}_{{\rm{0}}}}$ is the
nondimensional spin of the BH.  There are a total of 77 sets of data for different $(q,{a^{{\rm{BH}}}_{{\rm{0}}}},\Lambda,{m_{\rm{f}}},{a_{\rm{f}}})$ 
which are also collected in \cite{2019PhRvL.123d1102Z} (see its table II), with $q$ in the range $[2,7]$, $a^{{\rm{BH}}}_{{\rm{0}}}$ in the range $[-0.5,0.75]$, ${{\Lambda}}$ in the range [142, 2327],  and ${a_{\rm{f}}}$ in the range [0.32,0.91]. 
The equation (\ref{eq:af}) and (\ref{eq:BH-NS}) used to fit the data.

For NS-NS merger,  I adopt the numerical data collected in \cite{2019MNRAS.489L..91C} (table 3).  And again, the assumtion $a_{\rm{NS}}=0$ is adopted, hence $\chi=0$. There are a total of 21 sets of data for different $(q,M,\tilde{{\Lambda}},m_{\rm{f}},{a_{\rm{f}}})$, with $q$ in [1,1.75], $M$ in [2.70,3.20] in the unit of a  solar mass, $\tilde{{\Lambda}}$ in [127,1106] and ${a_{\rm{f}}}$ in [0.544,0.828]. 
Similarly, equation (\ref{eq:af}) and (\ref{eq:NS-NS}) are used to  fit the data.

\subsection{Fitting Results}	
\begin{figure}
	\centering
	\includegraphics[width=1\textwidth, angle=0]{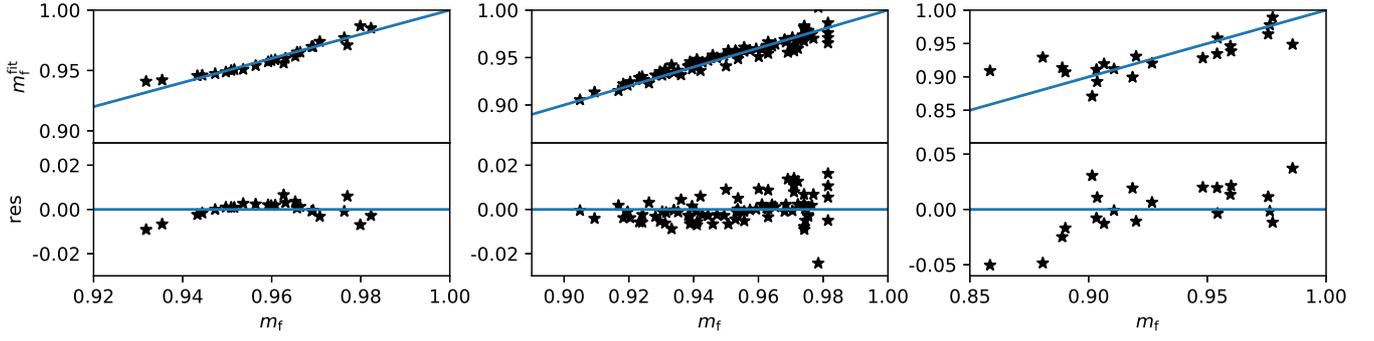}
	\caption{The plots of fitting result for final mass, from left to right corresponding to the cases of  BH-BH merger, BH-NS merger and NS-NS merger respectively. The upper panel is the comparison of the data and the fitting values. The lower panel is the residuals of the fit.}
	\label{fig:1}
\end{figure}

\begin{figure}
	\centering
	\includegraphics[width=1\textwidth, angle=0]{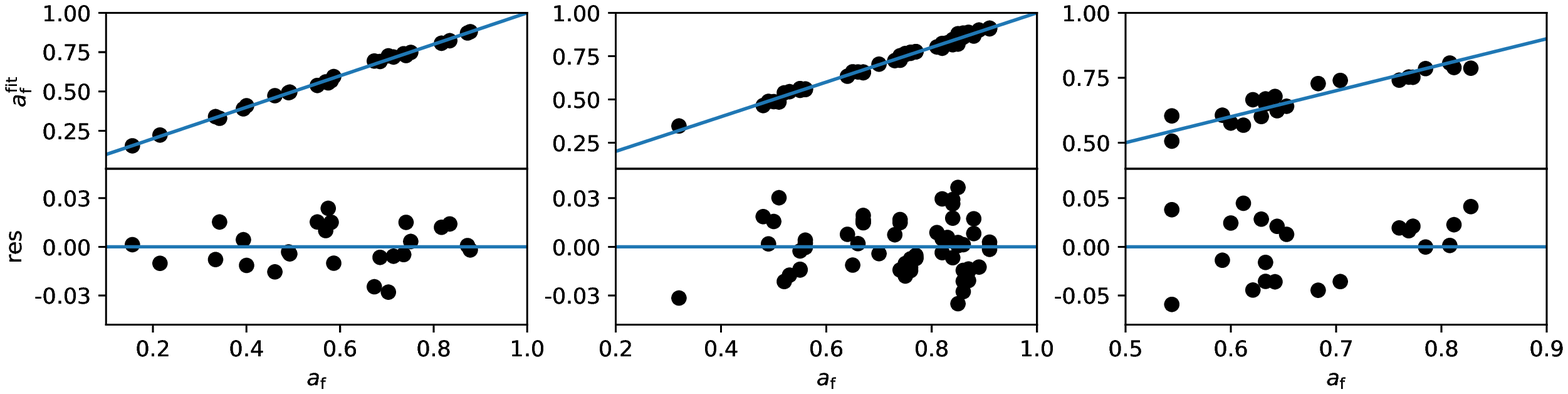}
	\caption{The plots of fitting result for final spin, from left to right corresponding to the cases of  BH-BH merger, BH-NS merger and NS-NS merger respectively. The upper panel is the comparison of the data and the fitting values. The lower panel is the residuals of the fit.}
	\label{fig:2}
\end{figure}

The fitting parameters are shown in table 1 for $m_{\rm{f}}$ and in table 2 for $a_{\rm{f}}$, respectively.  The comparison of $m_{\rm{f}}$ and the predicted $m_{\rm{f}}$  of the fit, as well as the residuals are plotted in figure \ref{fig:1}.  And the plots for $a_{\rm{f}}$  are shown in figure \ref{fig:2}.
One can see that the models for both $m_{\rm{f}}$ and $a_{\rm{f}}$ fit the data well overall, with small average absolute errors for the case of BH-BH and BH-NS mergers. For NS-NS merger, the fitting error is larger. However, this is within the acceptable range for astrophysical applications. And larger uniform samples for NS-NS merger are urgently needed to extend the presented results in the future.

From the fitting results, one sees that the model for $a_{\rm{f}}$ does have clear physical meaning as described in section 2. For non-spinning binaries one has $a_{\rm{f}}({\rm{BH-BH}}) \lesssim 0.82$, $a_{\rm{f}}({\rm{BH-NS}}) \lesssim 0.79$ and $a_{\rm{f}}({\rm{NS-NS}}) \lesssim 0.84$. It indicates that extremely fast rotating BH cannot be produced by the mergers of binary unless there is a very large initial spin $a_{1},a_{2} \sim 1$ and  a nearly equal mass $q \sim 1$. 
However, according to the current observation samples, it seems that the spins of the BHs in the merged binaries  are usually very small \citep{2019PhRvX...9c1040A}.

One also can understand the models for $m_{\rm{f}}$ somewhat naive according to the fitting results.  $m_{\rm{f}}$ is approximately equal to the initial mass minus the mass lost due to gravitational radiation and mass ejection i.e. $m_{\rm{f}} \approx 1-\Delta m_{\rm{rad}}-\Delta m_{\rm{ej}}$. This is straightforward for the
cases of BH-BH and BH-NS mergers, where $A_{\rm{m}} \approx 1 $ and $B_{\rm{m}}$, $C_{\rm{m}}$, $D_{\rm{m}}  <0$. The total mass lost is a combinational effect of the mass ratio, initial spin, tidal deformability and maybe other factors. However, the case of NS-NS merger is more complicated.

\begin{table}
    \centering
	\label{pm} 
	\caption{The fitting parameters for the final mass models, as well as the average absolute residual (AAE) and fractional error (FE).}
	\begin{tabular}{cccccccc}
		\hline
		\hline
		& A$_{\rm{m}}$     & B$_{\rm{m}}$     & C$_{\rm{m}}$      & D$_{\rm{m}}$    & AAE   &FE    \\
		\hline
		BH-BH & 0.988 & -0.610 & -0.042     & - & 0.0028 & 0.29\% \\
		BH-NS & 1.111 & -1.623 & -0.087 & -0.029 &0.0047 & 0.49\% \\
		NS-NS & 0.530 & 5.069   & 0.608 & -0.069 & 0.0180 & 1.96\%\\
		\hline
		\hline
	\end{tabular}
\end{table}

\begin{table}
	\centering
	\label{pa} 
	\caption{The fitting parameters for the final spin model, as well as the average absolute residual (AAE) and fractional error (FE).}
	\begin{tabular}{ccccccc}
		\hline
		\hline
		& A$_{a}$     & B$_{a}$     & C$_{a}$         & AAE   &FE    \\
		\hline
		BH-BH & 3.270 & 0.846 & 2.583   & 0.0088& 1.67\% \\
		BH-NS & 3.176 & 0.869 & 1.265  & 0.0101 & 1.42\% \\
		NS-NS & 3.346 & -        & 2.101 & 0.0275 & 4.23\%\\
		\hline
		\hline
	\end{tabular}
\end{table}

\section{Rotational Energy of the RBH}
The rotational energy $E_{\rm{rot}}$ of a BH is determined by its spin and mass, namely
\begin{equation}
E_{\mathrm{rot}}=f(a _{\rm{f}}) \frac{M_{\rm{f}}}{M_{\odot}} c^{2} \mathrm{erg}=1.8 \times 10^{54} f(a _{\rm{f}})\frac{M_{\rm{f}}}{M_{\odot}} \mathrm{erg},
\end{equation}
where $f(a_{\rm{f}})=1-\sqrt{(1+\sqrt{1-a_{\rm{f}}^{2}}) / 2}$.
	
Based on the model for final mass and spin from the previous section, one can then calculate the distribution of rotational energy
of the RBHs. 
For BH-BH merger, if binary BHs arise from the evolution of isolated binary stars, the primary star will form the more massive BH first, and the secondary star will form the less massive BH later \citep{2016Natur.534..512B}. Numerical simulation shows that the spin of the primary BH is very small, while the spin of the secondary BH can be large \citep{2018A&A...616A..28Q}.
Therefore, let's consider this situation here, where results in $\chi=(a_1)/{(1 + q)^2}$, $a_1$ is the spin of the secondary BH.
Assume a total mass of [20, 60] solar mass, $q$ in [1,7] and $a_1$ in [-1,1]. Surprisingly, the distribution of the rotation energy is very narrow within these parameters range which are considered, basically around $10^{54}$erg.

For binary NSs merger, let's consider $q$ in [1,2], $M$ in [2.7,3.2] and $\tilde{{\Lambda}}$ in [127,1106]. Interestingly, the rotational energy of the RBH is also in a narrow range, essentially on the order of $10^{53}$erg.  This is similar to the case of BH-BH merger but with an order of magnitude lower rotational energy.

\begin{figure}
	\centering
	\includegraphics[width=0.7\textwidth, angle=0]{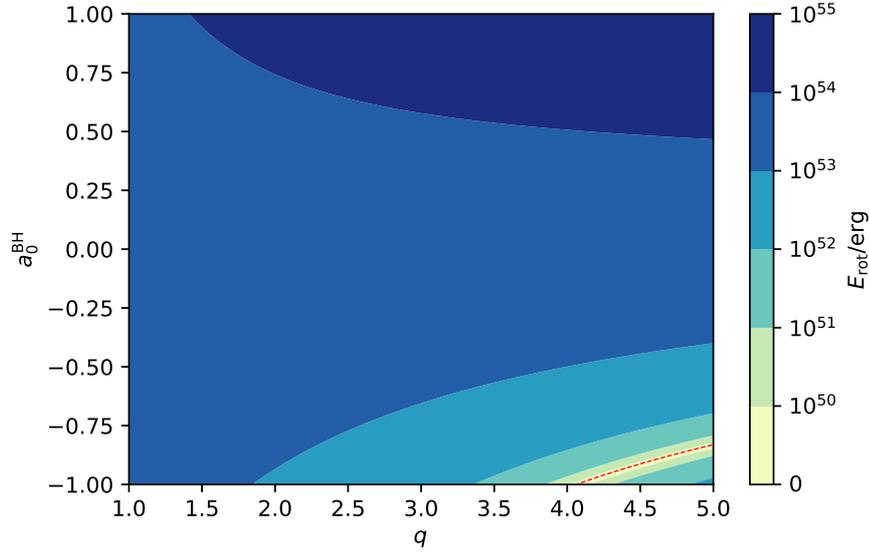}
	\caption{Contour map of the rotational energy of the RBH  as function of $(q,a^{\rm{BH}}_0)$ for BH-NS merger, by assuming ${\Lambda} =600$. It can be seen that the distribution of the rotational energy is not monotonic in $q>4$ region, and it is reasonable. Because $a_{\rm{f}}$ can change from positive through 0 to negative, however, the rotation energy is always non-negative. The red dash line in the figure is the region where the rotational energy is $\sim 0$ and $a_{\rm{f}} \sim 0$.  Above the line $a_{\rm{f}}>0$, and below the line $a_{\rm{f}}<0$.}
	\label{fig:3}
\end{figure}

Figure \ref{fig:3} shows the distribution of the rotational energy of the RBHs from the mergers of BH-NS  in the parameter space $(q,a^{\rm{BH}}_0)$. I confirm that the effect of ${\Lambda}$ is small, and it does not affect the result of the magnitude estimation for the rotational energy of the RBHs, thus a rough value ${\Lambda} =600$ is adopted. It can be seen that the rotational energy in most parameter spaces is distributed in the range $[10^{52},10^{54}]$erg.
In the moderate parameter space, the rotation energy is on the order of $10^{53}$erg. Even more interesting, in some narrow parameter spaces, the rotational energy can be as small as 0, that is, merging can produce schwarzschild BHs 
\footnote{  It should be pointed out that this depends on the extrapolation of the model to $a _{\rm{f}}\rightarrow 0$. Although the $a _{\rm{f}}$ only covers [0.32,0.91] in the fit, we believe that this conclusion is at least qualitatively reliable.}.
Therefore, the distribution of the rotational energy of the RBH from BH-NS merger can be in a wide range. This is vary different from the cases of the BH-BH and BH-NS mergers.

It can be seen that the distribution of the rotational energy of the RBHs produced by different types of mergers of compact binary has its own characteristics, and the RBHs with different rotational energy might cause their electromagnetic counterparts to be different.
Recently, LIGO detected a candidate of BH-NS coalescence with $q$ in the range of $[2.14, 5.01]$ \citep{2019arXiv191200375Y}.
Interestingly, there is a short gamma-ray signal that might be associated with it, and with an energy of $10^{48}$erg \citep{2019arXiv191200375Y}.
Considering the energy conversion efficiency around $10^{-4}-10^{-2}$, the required energy budget would be on the order of $10^{50}-10^{52}$erg. 
Analogously with the GRB model, assuming the source of the energy is from the rotational energy of the RBH extracted by BZ mechanism, then the BH-NS merger is more likely, if this GW event and the association are real. Such a rotational  energy of RBH, as seen in figure \ref{fig:3} , would require the component BH have a  relatively large inverse spin which has yet to be tested by gravitational wave detection. If such mutual verification is possible through gravitational waves detection, then it had the potential to help identify the electromagnetic counterparts of the merged events, especially interesting for the events without precise localization.

\section{Conclusion}
In this paper, I develops unified models for calculating the final mass and spin from  compact binary mergers with a good accuracy,  which is convenient for the astrophysical applications.  Compared with the previous models, our models is very simple but meanwhile the accuracy of the model is good to a certain extent. Furthermore,  I systematically studied the distribution of the rotation energy of RBHs, which is relevant to the the electromagnetic counterparts of the compact binary mergers. I find  that the distribution of the rotational energy of the RBHs produced by different types of mergers of compact binary has its own characteristics, which might help identify the electromagnetic counterparts of the merged events.

\section{Acknowledgments}
I would like to thank the referee for the very careful and helpful comments and suggestions that have allowed
me to improve the presentation of this manuscript.
This work is supported by the China Postdoctoral Science Foundation and  the Fundamental Research Funds for the Central Universities (NO. WK2030000019).

\section{Data Availability Statements}	
The data underlying this article will be shared on reasonable request
to the corresponding author.

\clearpage	
\onecolumn

\end{document}